\documentclass[twocolumn]{aastex63}
\usepackage{amsmath}
\usepackage{amsfonts}
\usepackage{amssymb}
\usepackage{bm}
\usepackage{color}
\usepackage{graphicx}
\usepackage{natbib}
\usepackage[utf8]{inputenc}

\graphicspath{ {./figures/} }

\usepackage{tikz}
\usetikzlibrary{positioning,calc,angles,quotes,math}
\usepackage{tikz-3dplot}

\shorttitle{The Stationary Point}
\shortauthors{Van Kooten et al.}

\received{September 30, 2024}
\revised{January 17, 2025}
\accepted{February 1, 2025}
\submitjournal{\apj}

\newcommand{\Rsun}{R\ensuremath{_\odot}}
\renewcommand{\deg}{\ensuremath{^\circ}}
\newcommand{\kms}{km~s$^{-1}$}
\renewcommand{\vec}[1]{\ensuremath{\boldsymbol{\mathbf{#1}}}}

\begin{document}

\title{The Stationary Point: A New Method for Solar Wind Speed Measurements from a Moving Vantage Point}

\author[0000-0002-4472-8517]{Samuel J. Van Kooten}
\affil{Southwest Research Institute, 1301 Walnut St. Suite 400, Boulder CO 80302}

\author[0000-0002-7164-2786]{Craig E. Deforest}
\affil{Southwest Research Institute, 1301 Walnut St. Suite 400, Boulder CO 80302}

\author[0000-0001-8480-947X]{Guillermo Stenborg}
\affil{The Johns Hopkins University Applied Physics Laboratory, Laurel, MD 20723, USA}

\author[0000-0002-2390-5932]{Kenny N. Kenny}
\affiliation{University of Colorado, Boulder, Department of Astrophysical and Planetary Sciences \\
2000 Colorado Ave, Boulder, CO 80305, USA}
\affil{Southwest Research Institute, 1301 Walnut St. Suite 400, Boulder CO 80302}

\correspondingauthor{Samuel Van Kooten}
\email{samuel.vankooten@swri.org}

\begin{abstract}
	The WISPR imager on Parker Solar Probe provides a unique view the young
	solar wind, flying through solar wind structures at high speed. It is
	of interest to use WISPR image sequences to measure the velocity of
	both large features (such as CMEs) and the background, ambient wind.
	However, WISPR's close-up, rapidly-moving perspective makes the usual
	methods for measuring velocities from images difficult or impossible to
	apply, as most apparent motion through the image is due to the motion
	or rotation of the imager. In this work, we propose a new method of
	looking for features at the ``stationary point''---a direction from
	which some plasma parcels appear to approach the spacecraft, remaining
	at a constant direction in the image sequence. This direction is a
	function of the plasma's radial velocity, the encounter geometry, and
	the spacecraft velocity, allowing the former two to be inferred. We
	demonstrate the technique with forward-modeled images, and we apply it
	to WISPR observations, inferring the speed and trajectory of a
	particular density feature. This method promises to enable speed
	measurements of the young solar wind in an important acceleration
	region, from a close-up perspective and at latitudes well outside the
	PSP orbital plane. And while we present this method in a solar wind
	context, it is broadly applicable to any situation of a moving
	viewpoint traveling through an expanding cloud of features.
\end{abstract}

\keywords{Solar wind (1534); Solar corona (1483); Solar coronal transients (312)}

\section{Introduction}
\label{sec:intro}

The solar wind is a continual outflow of heated plasma from the Sun which
fills the heliosphere, and the cause of its acceleration is an open question
of great interest in the field \citep{Viall2020} for which many competing
mechanisms have been proposed \citep[see, e.g., discussions
in][]{Cranmer2017,Cranmer2019}. The main difficulty in resolving the question is
the lack of strong observational constraints that are able to differentiate
between the proposed mechanisms (though recent observations have shed new light,
e.g. \citealp{Bale2023}). One type of observation that can contribute is a
measurement of solar wind flow speeds at a range of distances from the Sun, as
this will constrain where and to what degree acceleration is occurring. To this
end, we propose a new method for measuring wind flow speeds with the Wide-field
Imager for Parker Solar PRobe (WISPR; \citealp{Vourlidas2016}) imager on board
Parker Solar Probe (PSP; \citealp{Fox2016}). PSP's extreme proximity to the Sun
($\sim10$~\Rsun\ near perihelion in the latter portion of the planned mission)
and WISPR's field of view reaching to within 13\deg\ of the Sun allow WISPR to
probe the solar wind very close to the Sun---as low as a few \Rsun---in a region
where the majority of the wind's acceleration has been seen to occur
\citep[e.g.][]{Wexler2020}.

It has long been known that inhomogeneities pervade the outer corona and the
young solar wind \citep[e.g.][]{Sheeley1997,Viall2015,DeForest2018}, providing
visual tracers of the wind's motion, and tracking these density features
allows a remote measurement of the wind's speed. Such tracking applied to WISPR
requires new or adapted techniques, as traditional methods of flow tracking and
speed inference are difficult or impossible to apply. This is due primarily to
PSP's rapid motion and close proximity to the young solar wind, as well as the
constant rotation of WISPR's field of view (as PSP rotates to maintain heat
shield alignment). These effects mean that the true motion of a feature
of interest drives only a portion of its motion through the image plane, with
parallax and field-of-view rotation causing the remainder. We therefore develop
a new technique, in which a feature's apparent motion through the image plane,
combined with a full consideration of the observing geometry, allows a feature's
speed and trajectory to be inferred. While developed for WISPR, this technique
is in no way limited to this one application, but could be used in similar
situations in which one's viewpoint is moving steadily through a cloud of
radially-expanding features.

Other successful efforts to extract plasma positions and velocities from WISPR
images exist. \citet{Liewer2019,Liewer2020} developed a technique for
determining the trajectory of density features in WISPR images that make use of
PSP's rapid motion near perihelion. Using the multiple views of the feature over
several hours, its direction and velocity can be found by tracking the feature's
changing location in the images and fitting this sequence with an analytic
expression relating image-plane motion to motion in a heliocentric coordinate
frame. A modification of this technique \citep{Liewer2022,Liewer2023} can be
used to determine the coordinates of ray-like features by tracking points
distributed along the ray over the course of a few hours and again fitting the
motion to the analytic expressions. A method proposed by \citet{Kenny2023} will
extract the locations of ray-like structures which pass over or under WISPR by
using only the apparent motion of the ray through the image plane. This
technique promises to easily scale to track all rays seen by WISPR. When a
feature can be identified with WISPR and another imager at a different location,
its location can be triangulated from these separate viewpoints
\citep{Liewer2021}. (Such triangulation applied in solar physics dates back at
least as far as the early days of the Solar TErrestrial RElations Observatory
mission; \citealp{Thompson2009}.) Similarly, \citet{Braga2021} extended the
method of \citet{Liewer2020} to use a second viewpoint, allowing them to relax
the assumption that the feature is moving radially at constant velocity. As a
final example, \citet{Nistico2020} explore the use of feature brightness to
allow features to be located solely from WISPR observations. The feature's
distance from the Sun and its location relative to the Thomson sphere affect its
observed brightness, so the observed variation of brightness with time adds
additional information that can be fit alongside the feature's observed
elongation to determine its location and speed.

In this paper, we introduce our method, which we call the ``stationary point
method.'' We develop the approach in three phases, considering first the case of
PSP moving in a straight line at constant velocity, with plasma features in
the orbital plane. We extend to a realistic PSP trajectory, and then we build
further to the 3D case, where the plasma being observed is outside the orbital
plane. Each stage is demonstrated with synthetic images from a forward model.
Having shown that speeds and trajectories can be accurately inferred in the 3D
case, we next apply the method to a real plasma feature seen by WISPR,
demonstrating its viability in the real world. We leave to planned future work
the task of measuring many such features across the WISPR data catalog.

\section{The Stationary Point method}
\label{sec:method}

\subsection{Introduction by analogy}

The \textit{stationary point} concept can be introduced quickly by analogy to
driving a car during a snow storm (a familiar situation for some readers---and
these authors). While the snow is falling downward, the car's forward velocity
makes it appear in the car's reference frame that the snow is traveling with
some horizontal velocity (the negative of the car's velocity), and the snow
appears to
approach the car from a direction somewhere between horizontal and vertical.
If the car speeds up, the snow's apparent approach direction becomes closer to
horizontal. If the car stops, the snow appears to the driver to be falling
vertically. If the car maintains its speed but the snow's (true) vertical
velocity were somehow increased significantly, its apparent approach direction
would be closer to vertical. It is clear that this apparent approach direction
is a function of the snow's velocity, the car's velocity, and the geometry of
the situation (i.e. the snow is falling down with a uniform velocity, and the
car is maintaining straight, horizontal motion). Since the latter two are known
by the driver, a measurement of this approach direction can allow the snow's
velocity to be inferred.

This concept applies just as well to PSP flying through a cloud of
radially-outflowing density enhancements in the solar wind. We focus
specifically on those density features which eventually collide with the
spacecraft (or rather, through which the spacecraft travels) or which pass
directly over or under PSP.

\subsection{Straight-line spacecraft motion}
\label{sec:straight-line}
\subsubsection{Geometry}

\begin{figure}[t]
	\centering
	\includegraphics{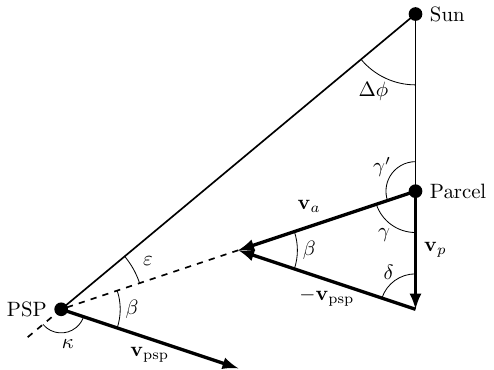}
	\caption{Diagram of stationary point geometry.}
	\label{fig:stationary-point}
\end{figure}

\begin{figure}[t]
	\centering
	\includegraphics{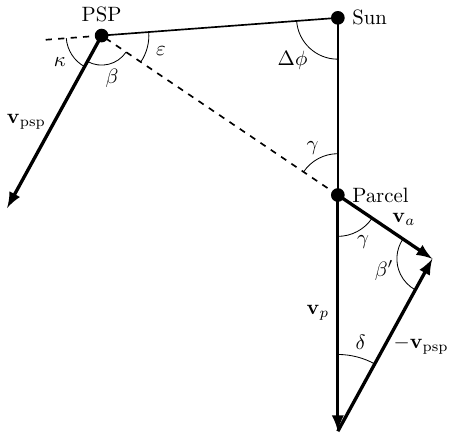}
	\caption{Alternative geometry for the case in which the parcel is growing more distant from the spacecraft.}
	\label{fig:stationary-point-away}
\end{figure}

In a first, simplified case in which PSP is traveling in a straight line with a
constant velocity and the plasma being observed is in the orbital plane,
collision-course plasma parcels appear to approach PSP from a fixed direction,
because in PSP's rest frame such parcels are directly approaching PPS, and
therefore the parcels remain on the same line of sight and appear at the same position
in the WISPR image plane throughout their approach. This is the origin of the
term ``stationary point,'' referring to the angular position at which these
parcels appear stationary (though growing larger)---as opposed to the majority of
parcels, on non-collision-course trajectories, which move in apparent position
in the imaging plane. An alternative case is also possible, in which the parcel
is growing more distant along the line of sight, rather than approaching PSP.
This corresponds to a parcel which previously ``collided'' with PSP (from
behind, if it is now being observed by the forward-facing camera), and such
parcels also appear at a constant angular position.

We develop this geometry further with the aid of
Figures~\ref{fig:stationary-point} and \ref{fig:stationary-point-away} for the
``parcel approaching'' and ``parcel retreating'' cases, respectively, focusing
first on the ``approaching'' case. A plasma parcel moves radially out from the
Sun at constant velocity $v_p$, and PSP moves in a straight line at constant
velocity $v_\textrm{psp}$. In the spacecraft frame of reference, PSP is
stationary and the parcel moves with an apparent velocity $\vec{v}_a = \vec{v}_p
- \vec{v}_\textrm{psp}$. In the two cases shown, $\vec{v}_a$ is pointed directly
toward or away from PSP, indicating a future or past collision, and also
indicating the parcel will always be seen in the same direction---that of the
stationary point, a constant angle $\beta$ relative to the forward direction of
the spacecraft (defined as the angle between $\vec{v}_\textrm{psp}$ and
$\vec{v}_a$'s continuation to PSP---the latter vector is equivalent to the line
of sight along which the parcel is seen).

We also mark $\beta$ where it appears again in the triangle of velocity
vectors as well as the angles $\varepsilon$, the elongation from the Sun at
which the parcel is seen; $\kappa$, the angle between $\vec{v}_\textrm{psp}$ and
the continuation of the Sun--PSP line; $\Delta\phi$, the longitudinal separation
between the parcel and PSP; $\delta$, the angle between $\vec{v}_p$ and
$\vec{v}_\textrm{psp}$; $\gamma$, the angle between $\vec{v}_p$ and $\vec{v}_a$;
$\gamma^\prime = 180\deg - \gamma$, and $\beta^\prime = 180\deg - \beta$. Some
of these angles will not be used until Section \ref{sec:3D}.

\begin{figure}[t]
	\centering
	\includegraphics{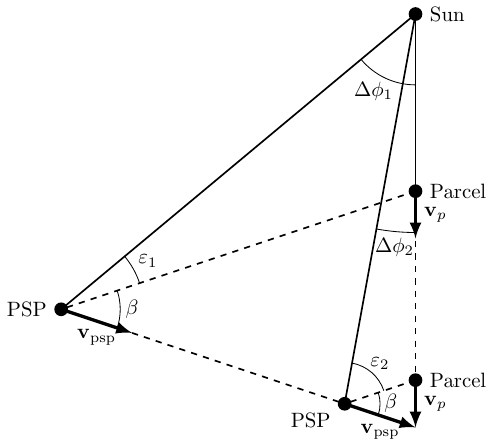}
	\caption{The situation just before collision. Shown are the same PSP and
	parcel positions as in Figure~\ref{fig:stationary-point}, as well as later
	positions when the two are much closer together. $\beta$ is the same across
	the two cases by the nature of the ``collision course'' geometry, but
	$\varepsilon$ has increased and $\Delta\phi$ has decreased.}
	\label{fig:stationary-point-just-before-collision}
\end{figure}

To produce an expression relating the stationary point location to the plasma
velocity, we look at the Sun--parcel--spacecraft triangle in
Figure~\ref{fig:stationary-point} and write
\begin{align}
	180\deg &= (180\deg - \gamma) + \varepsilon + \Delta\phi \\
	\Rightarrow \gamma &= \epsilon + \Delta\phi.
\end{align}
In the triangle created by the parcel's velocity vectors, we use the law of
sines:
\begin{align}
	\label{eqn:law-of-sines}
	\frac{v_\textrm{psp}}{\sin\gamma} &= \frac{v_p}{\sin\beta} \\
	v_p &= \frac{v_\textrm{psp} \sin\beta}{\sin\left(\varepsilon + \Delta\phi\right)}
	\label{eqn:constr1}
\end{align}
This contains $\Delta\phi$, the longitudinal separation between
spacecraft and parcel, which is not known. However, this can be
simplified by considering for now only the moments just before
or just after collision, illustrated in
Figure~\ref{fig:stationary-point-just-before-collision}, when the longitudinal
separation between the parcel and PSP
approaches zero and therefore $\Delta\phi \ll \varepsilon$, so
\begin{align}
	v_p &= v_\textrm{psp}\frac{\sin\beta}{\sin\varepsilon}.
	\label{eq:stationary-point}
\end{align}
In these moments immediately before or after collision, the parcel velocity can
thus be determined from the known spacecraft velocity and the stationary point's
measured location measured relative to two reference points: the Sun (for
$\varepsilon$) and the spacecraft's velocity direction (for $\beta$). The latter
direction is fixed in this scenario, and so $\beta$ can be converted from an
angle relative to any other convenient reference direction---for example, the
center point of the camera field of view.

For the ``parcel retreating'' case of Figure~\ref{fig:stationary-point-away},
(i.e. $\vec{v}_a$ pointing directly away from the spacecraft instead of directly
toward it), in Equation~\ref{eqn:law-of-sines} $\sin\beta$ becomes
$\sin\beta^\prime = \sin(180\deg-\beta) = \sin\beta$, and the expression for
$\sin\gamma$ becomes $\sin(180\deg-\varepsilon-\Delta\phi) =
\sin(\varepsilon+\Delta\phi)$, so Equations~\ref{eqn:constr1} and
\ref{eq:stationary-point} are unchanged. The two cases will, however, become
distinct in the 3D geometry of Section~\ref{sec:3D}.

The division between the ``parcel approaching'' and ``retreating'' cases occurs
when $\Delta\phi$ is such that $\vec{v}_p$ is parallel to
$\vec{v}_\textrm{psp}$. In this case, for the parcel to remain on the same line
of sight, the two velocities---which match in direction---must also match in
magnitude, and so the plasma parcel is at a fixed location in the spacecraft
frame, neither approaching nor retreating from PSP. This dividing value is
$\Delta\phi = \kappa$, where $\kappa = 180\deg - \beta - \varepsilon$. For
$\Delta\phi < \kappa$, $\vec{v}_p$ has a component toward PSP, and for
$\Delta\phi > \kappa$ it has a component away from PSP, making these two regimes
the ``approaching'' and ``retreating'' cases, respectively.

\begin{figure}[t]
	\centering
	\includegraphics[width=\columnwidth]{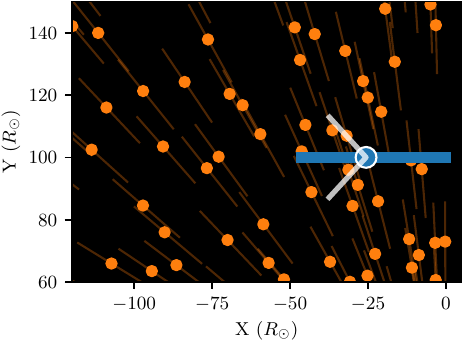} \\
	\vspace{.3cm}
	\includegraphics[width=\columnwidth]{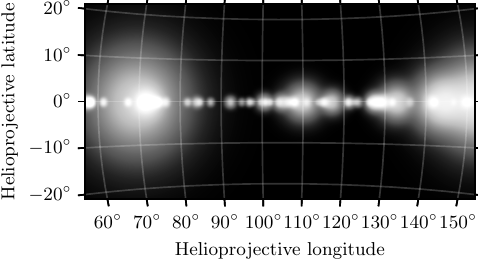} \\
	\vspace{.3cm}
	\includegraphics[width=\columnwidth]{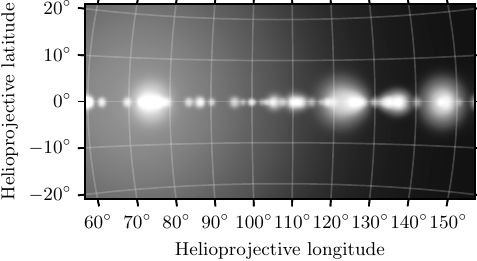}
	\caption{Top: an overhead view of our model setup, with the spacecraft (blue
	dot) traveling to the left in a straight line (blue) and its field of view
	indicated by the thin white lines. A number of plasma parcels (orange)
	travel radially out from the Sun, with their positions over the entire time
	range indicated by orange lines. Center and bottom: two sample synthesized
	images from this model, at the same point in time as the top panel and a
	short time later. Of note is a large parcel on the left, which is rapidly
	growing in the field of view as it approaches the spacecraft on a collision
	course. (The left
	edge of the images corresponds to the bottom portion of the marked
	field of view.)}
	\label{fig:jmap-model-straight-line-overhead}
\end{figure}

\begin{figure*}[t]
	\centering
	\includegraphics{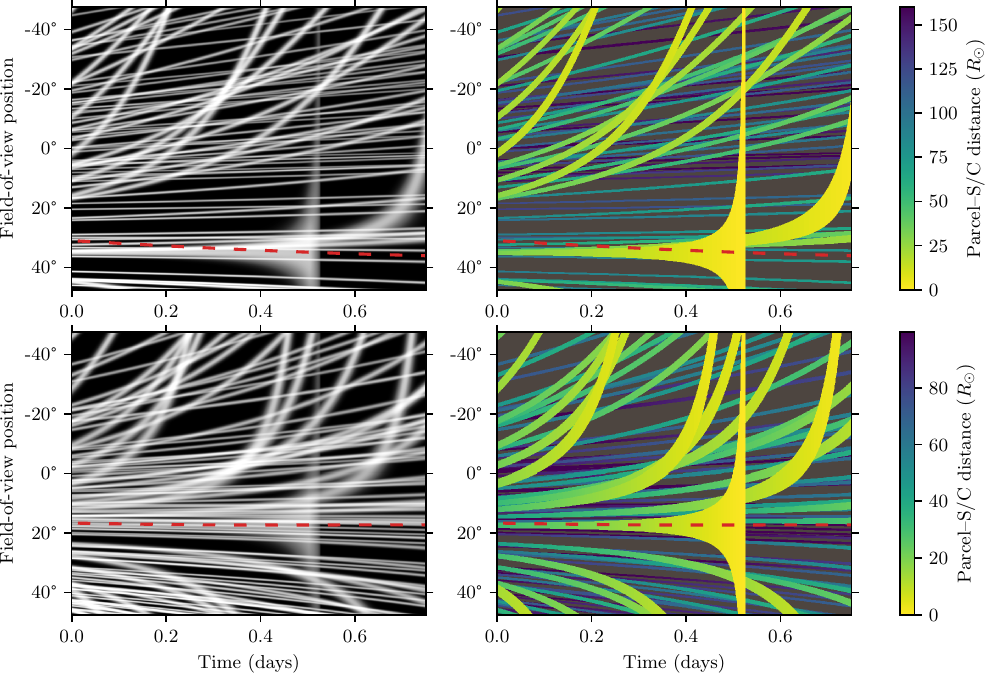}
	\caption{Synthetic time--distance plots produced by straight-line spacecraft motion
	through a cloud of plasma parcels (as shown in
	Figure~\ref{fig:jmap-model-straight-line-overhead}). The parcels move
	radially-out from the Sun at 300~\kms\ (top) or 150~\kms\ (bottom) and the
	spacecraft moves at 500~\kms. The left column shows simulated white-light
	flux, while the right column shows the distance from the spacecraft to the
	nearest parcel along each line of sight, to illustrate which parcels are
	passing close to the spacecraft. The red dashed line in each plot shows the
	expected stationary point position for a hypothetical parcel colliding with
	PSP at that moment,	as given by Equation~\ref{eq:stationary-point}.
	Field-of-view position is along the 0\deg\ latitude line of
	Figure~\ref{fig:jmap-model-straight-line-overhead} and corresponds to
	$\beta$ in Equation~\ref{eq:stationary-point}, with positive values closer
	to the Sun.}
	\label{fig:jmap-model-straight-line-jmaps}
\end{figure*}

\subsubsection{Model demonstration}

We demonstrate the stationary-point phenomenon with a model of a spacecraft
flying in a straight line through a cloud of plasma parcels, shown in a top-down
view in Figure~\ref{fig:jmap-model-straight-line-overhead}. A number of
spherical plasma parcels of uniform size travel radially-out from the Sun, all
at the same constant speed, in random directions and with random release times.
For clarity and expediency, these parcels only exist in the spacecraft orbital
plane (corresponding to the plane of Figure~\ref{fig:stationary-point}).

From this setup we synthesize images by casting rays from each image pixel and
summing the flux contributions of each parcel the rays intersect, assuming each
parcel has an intensity profile that is Gaussian with respect to the impact
parameter, and imposing a $1/r^2$ falloff with the Sun--parcel distance, as well
as the effects of expansion with increasing $r$ and white-light Thomson
scattering (though what is important here is the geometry of where a parcel is
seen, not how bright it is). Two sample images are shown in
Figure~\ref{fig:jmap-model-straight-line-overhead}. We then generate
time--distance plots (sometimes called ``J-maps''), shown in
Figure~\ref{fig:jmap-model-straight-line-jmaps} for two different plasma
velocities (i.e., two different model runs). We also show plots indicating the
distance to each feature seen in the time--distance plots, to aid in
discriminating foreground and background objects. The time--distance plots
represent a single strip extracted from each image and stacked to produce the
\textit{time} axis. The strip we extract is along the projection into the image
plane of the plane containing the plasma parcels, the Sun and the spacecraft.
(This projection may in principle be curved, depending on the camera pointing
and projection.) In this case, the strip is the line of 0\deg\ latitude, and the
\textit{distance} axis of the time--distance plots therefore represents
horizontal position in the image plane. This emphasis on slicing along the
projected orbital plane, as opposed to simply taking an arbitrary row of pixels
(though in this simple example the projected plane is in fact a row of pixels),
is because parcels in the orbital plane remain on this strip through the entire
image sequence, whereas parcels outside the orbital plane move in two dimensions
through the image plane and therefore cannot be represented on a time--distance
plot with a single \textit{distance} axis. This will become important in
Section~\ref{sec:curved}, when the projected plane is no longer a straight line,
and this restriction will be lifted in Section~\ref{sec:3D}.

It is readily apparent in these time--distance plots that there is a small range
of field-of-view positions at which we see horizontal motion paths, indicating a
feature remaining at a fixed angular position in the field of view, and it is
this location that we call the stationary point. Other streaks tend to curve
away from the stationary point. The apparent stationary point location aligns
well with the expected location given by Equation~\ref{eq:stationary-point},
indicated by the dashed red lines in each plot (which we compute with knowledge
of $v_p$ for this simulation). These lines indicate where a hypothetical plasma
parcel would appear if it were just about to collide with PSP at that moment
(and so the $\Delta\phi\ll\varepsilon$ approximation is valid). The expected
stationary point location varies with time---this reflects the changing location
of the Sun relative to PSP, which affects the approach direction of the
radial-out plasma parcels and changes the relationship between $\beta$ and
$\varepsilon$, both of which appear in Equation~\ref{eq:stationary-point}.
(Recall that the parcel appears at a fixed $\beta$, a constant angle relative to
the fixed PSP velocity direction, whereas $\varepsilon$ is that same angle
measured relative to a \emph{variable} reference point, the direction to the
Sun.)

Each model run includes one parcel placed to collide exactly with the spacecraft
at approximately 0.55~days. This parcel is seen to approach from a fixed
direction (being $\beta=35\deg,17\deg$; $\varepsilon=73\deg,91\deg$ at collision
for the upper and lower rows of Figure~\ref{fig:jmap-model-straight-line-jmaps},
respectively) and steadily grow larger and larger, before filling the field of
view as it washes over the spacecraft.

By comparison with the distance maps also
in Figure~\ref{fig:jmap-model-straight-line-jmaps}, we see that the streaks
which curve significantly correspond to parcels passing very close to the
spacecraft. These are ``near-miss'' parcels---ones which do not collide with the
spacecraft, but get very close to doing so. When they are further away, they
appear almost identical to true collision-course parcels and approach the
spacecraft from a constant angular position very close to the stationary point.
But as they approach and then miss the spacecraft, they are seen at a
rapidly-changing angular position as they move from in front of the spacecraft
to the side.

Comparing the expected and observed stationary points across these plots reveals
a few insights. First, the stationary point location varies as the parcel
velocity (the wind speed) varies between the two rows of
Figure~\ref{fig:jmap-model-straight-line-jmaps}, showing how a measurement of
the observed stationary point location can be used to measure the parcel
velocity. Second, as mentioned earlier, it can be seen that the expected
stationary point location varies slightly with time, due to the changing
relationship between $\varepsilon$ and $\beta$. Third, in some cases
(particularly the exact-collision parcel in the top row), some parcels that
appear stationary and which coincide with the expected stationary point when
they are close to the spacecraft are further from the expected stationary point
when they are more distant. This is again because the expected stationary point
location varies with time. If we had full knowledge of $\Delta\phi(t)$ for a
given parcel (which we do not, with real observations), plotting the expected
stationary point with time for that parcel using Equation~\ref{eqn:constr1}
would produce a horizontal line to match the observed horizontal motion track
(that is, the constant $\beta$ that characterizes this ``collision course''
encounter).

It is important to note that in this example, there is a single stationary point
(though with slight variation with time) because all the plasma parcels move
with uniform velocity. If there were a distribution of plasma velocities, there
would be a corresponding range of stationary point locations, one corresponding
to each actual plasma velocity, which would complicate observational
determination of the stationary point from a time--distance plot.

\subsection{Realistic spacecraft motion}
\label{sec:curved}

\begin{figure}[t]
	\centering
	\includegraphics[width=\columnwidth]{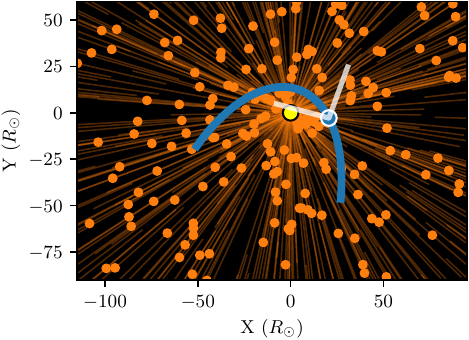} \\
	\vspace{.3cm}
	\includegraphics[width=\columnwidth]{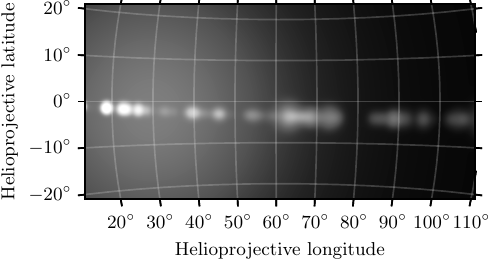}
	\caption{The same as Figure~\ref{fig:jmap-model-straight-line-overhead},
	but for a real PSP motion track (that of Encounter~13) and real WISPR
	pointing. The yellow dot marks the Sun. Note that the dots in
	the upper panel represent only locations, not to-scale sizes.}
	\label{fig:jmap-model-psp-overhead}
\end{figure}

\begin{figure}[t]
	\centering
	\includegraphics[width=0.99\columnwidth]{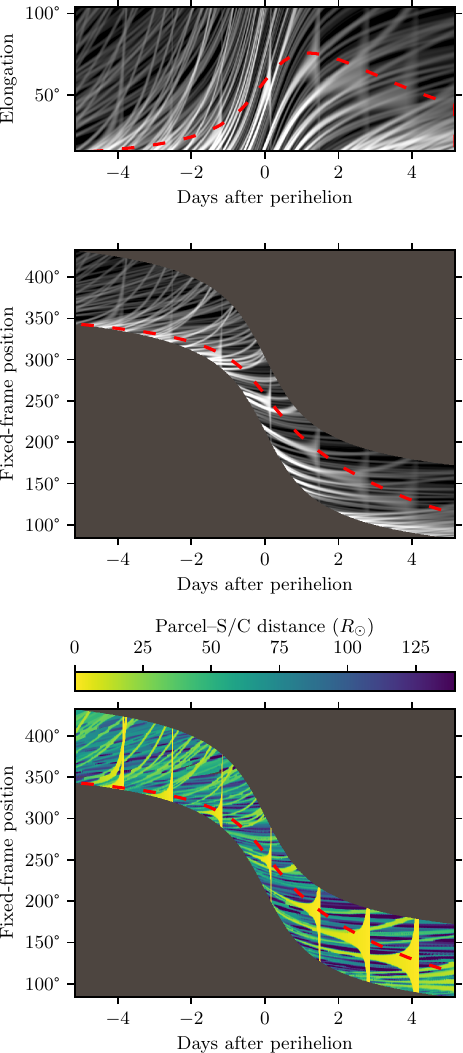}
	\caption{Time--distance plots for realistic PSP motion. Top: A traditional
	time--distance plot showing elongation (angular distance from the Sun).
	Middle: Our ``de-rotated'' plot, using a fixed-frame angular position as the
	distance axis. Bottom: A corresponding map of the distance from the
	spacecraft to the nearest parcel on each line of sight. The red dashed lines
	indicated the expected stationary point location. The uniform plasma speed
	is 100~\kms.}
	\label{fig:jmap-model-psp-jmaps}
\end{figure}

We now advance to a more realistic case, in which our model spacecraft follows
the trajectory of PSP (specifically, during Encounter~13). This model setup is
shown in Figure~\ref{fig:jmap-model-psp-overhead}. In this setup, the model
camera is pointed the same as WISPR: at a fixed elongation relative to the Sun.
(The fixed, Sun-relative pointing is driven entirely by the spacecraft
orientation, which maintains very careful heat shield alignment throughout each
encounter.) In the synthesized image shown in
Figure~\ref{fig:jmap-model-psp-overhead}, it can be seen that the plasma
parcels, which are placed only in the orbital plane, follow a line that is
tilted slightly relative to the helioprojective equator---this is due to the
inclination of the PSP orbital plane, and we again slice along the projected
orbital plane when building our time--distance plots.

This realistic case adds two complications over the straight-line case. First,
the spacecraft follows a very elliptical path with a rapidly-changing speed.
Second, the camera rotates rather quickly through the entire encounter. We
account for this rotation by modifying our time--distance plots so that the
``distance'' axis depicts the angular direction of each line of sight in a
fixed reference frame. In the overhead maps of
Figures~\ref{fig:jmap-model-straight-line-overhead} and
\ref{fig:jmap-model-psp-overhead}, this angle is that of a polar coordinate
system centered on the spacecraft, whose origin translates but does not rotate
to follow the spacecraft. (If the PSP orbital frame were aligned with the
celestial equator, this angle would be identical to right ascension.) We call
this presentation of the data a ``de-rotated'' time--distance plot, as it
subtracts out the rotation of the camera. We show this in
Figure~\ref{fig:jmap-model-psp-jmaps}, where we demonstrate an
elongation-versus-time plot, which is simply a strip along the orbital plane
taken from each image and stacked together, and our de-rotated plot, in which
each column from the first image is offset according to the pointing of the
camera for that image. We produce an ``expected stationary point'' location for
each time step by approximating the spacecraft velocity at that point as
constant in time and applying Equation~\ref{eq:stationary-point}. (This is
analogous to breaking the full time sequence into small windows of time in which
the spacecraft velocity is close to constant, and then treating each window
separately.) This location will describe where parcels are expected to appear in
the moments right before intercepting the spacecraft, but they will appear
further from the expected location at earlier points in time due to the
variation of the spacecraft velocity from the linearized approximation. In this
model run we include a number of ``direct-hit'' plasma parcels evenly spaced in
time, which can be seen as the features that grow rapidly in size before
disappearing and which are centered on the ``expected stationary point'' line in
their last moments before disappearing. These features follow very curved
trajectories in the elongation plot in the days before they intercept the
spacecraft, because the camera's rotation moves them quickly across the image
plane. In the de-rotated plot of Figure~\ref{fig:jmap-model-psp-jmaps}, they
still show a slightly angled track due to the relative motion between the parcel
and the spacecraft, but they appear closer to horizontal lines---this de-rotated
presentation of the time--distance map more closely resembles the
straightforward appearance of the stationary point in the straight-line motion
case earlier.

Comparing Figure~\ref{fig:jmap-model-psp-jmaps} to
Figure~\ref{fig:jmap-model-straight-line-jmaps} from the
straight-line-spacecraft case, it can be seen that collision-course parcels
deviate from the expected stationary point location much more strongly when
they are far from collision, due to the fact that at earlier times they actually
are not on a collision course. It is only in the moment of collision and a small
number of hours prior, when PSP's orbit is well-approximated by a straight line
and constant speed, that the stationary-point geometry holds. This makes it
harder to constrain the stationary point's location in the time--distance plot,
and motivates our progression to the next section.

\subsection{The Three-Dimensional Case}
\label{sec:3D}
\subsubsection{Geometry}
\label{sec:3D-geometry}

\begin{figure}
	\centering
	\includegraphics{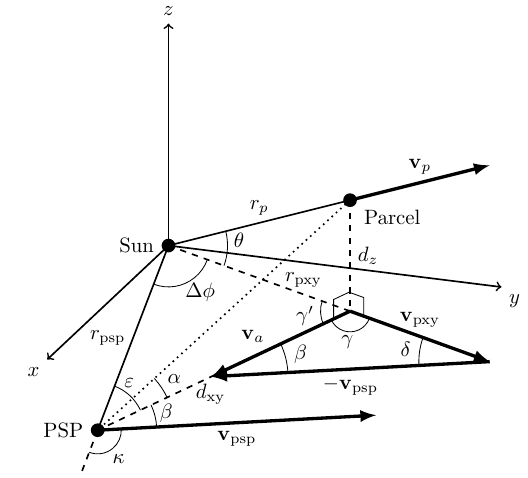}
	\caption{Diagram of 3D stationary point geometry. Everything is in the $x-y$
	plane except for the parcel, its velocity, and the angles $\alpha$ and $\theta$. The
	in-plane geometry is that of Figure~\ref{fig:stationary-point}, though the
	parcel location in that case has become here the in-plane projection of the
	parcel.}
	\label{fig:stationary-point-3d}
\end{figure}

We now extend the geometry to the third dimension, using the geometry of
Figure~\ref{fig:stationary-point-3d}. In this scenario, the plasma parcel is
outside the orbital plane of the spacecraft. Now, instead of the spacecraft and
parcel being on a collision course, the spacecraft is on a course to fly
directly under or over the parcel---in other words, it is on a collision course
with the projection of the parcel's position in the orbital plane. The same
approach of the previous sections, including approximating the spacecraft
velocity as constant for small windows of time, can therefore be used to measure
the in-plane projected velocity of the parcel, $\vec{v}_\textrm{pxy}$. However,
the 2D approach required limiting oneself to the moments right before or after
collision, when $\Delta \phi \ll \varepsilon$. In the 3D case, we can leverage
additional information to remove this limitation. Without it,
Equation~\ref{eqn:constr1} (with $v_p$ becoming $v_\textrm{pxy}$) produces a
curve in ($v_\textrm{pxy}$, $\Delta\phi$) space, relating these two unknown
quantities, and we use this as a first constraint ($C_1$). We now also use the
out-of-plane angle $\alpha$ at which the parcel is observed and that angle's
time derivative to add a second constraint ($C_2$), an additional curve which
produces a solution for ($v_\textrm{pxy}$, $\Delta\phi$).

Referring to Figure~\ref{fig:stationary-point-3d}, in the triangle of velocity
vectors, we use the law of cosines to write
\begin{equation}
	v_a = \pm \sqrt{v_\textrm{psp}^2 + v_\textrm{pxy}^2 - 2 v_\textrm{psp} v_\textrm{pxy} \cos\delta},
	\label{eqn:v_a}
\end{equation}
where $\delta$ can be written as $\delta = 180\deg - \beta - \varepsilon -
\Delta\phi$, by way of $\gamma^\prime$ and $\gamma$. Thus, $v_a$ is a function
of $v_\textrm{pxy}$ and $\Delta \phi$. The positive root describes the ``parcel
approaching'' case (for $\Delta\phi < \kappa$), while the negative root
describes the ``parcel retreating'' case (for $\Delta\phi > \kappa$). The
handling of these cases is discussed more later in this section.

Next, in the triangle containing the Sun, the spacecraft, and the in-plane
projection of the parcel, we use the law of sines to write
\begin{equation}
	d_\textrm{xy} = r_\textrm{psp} \sin \Delta \phi / \sin\gamma^\prime,
\end{equation}
where $\gamma^\prime = 180\deg - \varepsilon - \Delta\phi$ and therefore
$\sin\gamma^\prime = \sin\left(\varepsilon + \Delta\phi\right)$.

In the triangle containing the spacecraft, the parcel, and the parcel's in-plane
projection, we note that $d_z = d_\textrm{xy}\tan\alpha$. We compute
$r_\textrm{pxy}$ through the law of sines, $r_\textrm{pxy} = r_\textrm{psp}
\sin{\varepsilon} / \sin\gamma^\prime$. With this, we write $\theta =
\tan^{-1}\left(d_z / r_\textrm{pxy}\right)$ and then $v_{pz} =
v_\textrm{pxy}\tan\theta$ and $v_p = v_\textrm{pxy} / \cos\theta$.

Now introducing a time dependence, we write
\begin{align}
	\alpha(t) &= \tan^{-1}\left(\frac{d_z(t)}{d_\textrm{xy}(t)}\right), \\
	d_z(t) &= d_{z0} + v_{pz} t, \\
	d_\textrm{xy}(t) &= d_{xy0} - v_a t,
\end{align}
and we explore
\begin{align}
	\frac{d\alpha}{dt} &= \frac{d}{dt} \tan^{-1}\left(\frac{d_{z0} + v_{pz}t}{d_{xy0} - v_at}\right).
	\label{eqn:C2_before_subs}
\end{align}
Expanding and simplifying by setting $t=0$ (where this represents the time of
our observation), we reach (after substitutions)
\begin{equation}
	\frac{d\alpha}{dt} = \frac{v_a \tan\alpha + v_\textrm{pxy}\tan\theta}{d_{xy0} \left(1 + \tan^2\alpha\right)}.
	\label{eqn:constr2}
\end{equation}
Substituting further using the preceding equations, we arrive at the unwieldy
expression
\begin{multline}
	\frac{d\alpha}{dt} = \left[
		\pm \sqrt{v^2_\textrm{psp} + v^2_\textrm{pxy} + 2v_\textrm{psp}v_\textrm{pxy}\cos\left(\beta + \varepsilon + \Delta\phi\right)}\right. \\
		\cdot\frac{\tan\alpha\sin\left(\varepsilon + \Delta\phi\right)}{r_\textrm{psp}\sin\Delta\phi} \\
		\left.+ \frac{v_\textrm{pxy}\tan\alpha\sin\left(\varepsilon + \Delta\phi\right)}{r_\textrm{psp}\sin\varepsilon}\right] \\
		/ \left[1 + \tan^2\alpha\right],
		\label{eqn:expanded_C2}
\end{multline}
with the $\pm$ coming from $v_a$ to be discussed later. This expansion shows
that the two unknowns, $v_\textrm{pxy}$ and $\Delta\phi$, are related through
only the measured values of $\beta$, $\varepsilon$, $\alpha$ and $d\alpha/dt$,
plus the known values of $v_\textrm{psp}$ and $r_\textrm{psp}$, producing our
second constraint ($C_2$). The intersection of $C_1$ and $C_2$ produces a
solution for $v_\textrm{pxy}$ and $\Delta\phi$, with which all the quantities in
this section can be computed, producing a complete solution for the parcel's
location ($r_{p}$, $\Delta\phi$ and $\theta$) and assumed-radial velocity vector
($v_p$, $\Delta\phi$ and $\theta$)

\begin{figure}
	\centering
	\includegraphics{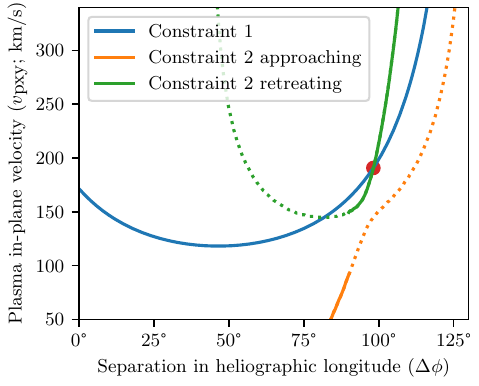}
	\caption{Constraints produced for the model setup of
	Figure~\ref{fig:3d-model-setup}. The solid portion for each of the two
	variants of the second constraint represent their domain of validity, and
	the dotted portions extend the curves outside that domain, to illustrate
	each more completely. The dot marks the
	intersection of the two constraints---a correct solution for $\Delta\phi,
	v_\textrm{pxy}$.}
	\label{fig:3d-model-intersects}
\end{figure}

In the case in which the parcel's in-plane projection is growing more distant
from the spacecraft, the in-plane geometry is that of
Figure~\ref{fig:stationary-point-away}. As shown in
Section~\ref{sec:straight-line}, the first constraint is unchanged. The
additional expressions in this section are similarly unchanged, since
$\sin\theta = \sin(180\deg-\theta)$, except that $v_a$ (Equation~\ref{eqn:v_a})
and the expanded expression in Equation~\ref{eqn:expanded_C2} take the negative
root rather than the positive. This produces two variants of the second
constraint, with mutually-exclusive domains divided by $\Delta\phi = \kappa$
which separates the ``approaching'' and ``retreating'' cases. The ``true'' $C_2$
is the union of these two variants, each restricted to its own domain.

We now briefly explore the behavior of the second constraint with the aid of
Figure~\ref{fig:3d-model-intersects}, plotting $C_1$ and $C_2$ for a model setup
we will present in Section~\ref{sec:3d-model-demo}---for now we only want an
example while discussing the general behavior of $C_2$. The plot shows each
$C_2$ variant in its entirety, with the solid portions of each curve indicating
each variant's domain of validity and the ``true'' $C_2$ curve being the union
of the two solid portions. At the dividing line of $\Delta\phi = \kappa =
180\deg - \beta - \varepsilon$, it is clear there is a large discontinuity in
this union. At
this point, the $\delta$ in Equation~\ref{eqn:v_a} is 0. Since, as discussed in
Section~\ref{sec:straight-line}, $\Delta\phi = \kappa$ represents a parcel (or
now its in-plane projection) moving parallel to PSP, with
$v_\textrm{pxy}=v_\textrm{psp}$ in order for the parcel to stay on the same line
of sight, $v_a$ becomes zero. This allows $v_a$ to be continuous as $\Delta\phi$
varies and crosses $\kappa$, changing the sign chosen for $v_a$. But this is
true only when the stationary point geometry holds, with the parcel remaining at
constant $\beta$. Since $C_1$ is an expression of that phenomenon, it
essentially produces the value of $v_p$ required for a parcel at a given
$\Delta\phi$ to appear at the observed stationary point $\beta$. On the other
hand, $C_2$ is an expression of how the out-of-plane angle $\alpha$ varies, and
it does not overlap $C_1$ except at their intersection, meaning the stationary
point geometry does not hold for most of the points on $C_2$. This means that at
$\Delta\phi=\kappa$, $C_2$'s $v_\textrm{pxy} \ne v_\textrm{psp}$, and so $v_a
\ne 0$ at the sign change, producing the discontinuity we see in the plot. As
the discontinuity is therefore always, if present, away from $C_1$, our solution
will never be at this discontinuity, while if the correct $\Delta\phi$ is
$\kappa$, $C_2$ will cross $C_1$ at that point, and so the stationary point
geometry will hold and $C_2$ will be continuous. It is important to note that
this line of thought implies that our values of $C_2$ are incorrect everywhere
except for its intersection with $C_1$ (and that its valid value at the
intersection is the only value we use). We derived $C_2$ from
Equation~\ref{eqn:C2_before_subs}, which is generally true, but we then
substituted values derived from the stationary point geometry. But for a given
$\Delta\phi$, the stationary point geometry requires the $v_\textrm{pxy}$ given
by $C_1$, which is not the $v_\textrm{pxy}$ given by $C_2$ outside the point of
intersection. This means that our final expression for $C_2$ is valid where it
intersects $C_1$ but inconsistent everywhere else, producing $C_2$'s
non-physical discontinuity.

\subsubsection{Model demonstration}
\label{sec:3d-model-demo}

\begin{figure*}
	\centering
	\includegraphics{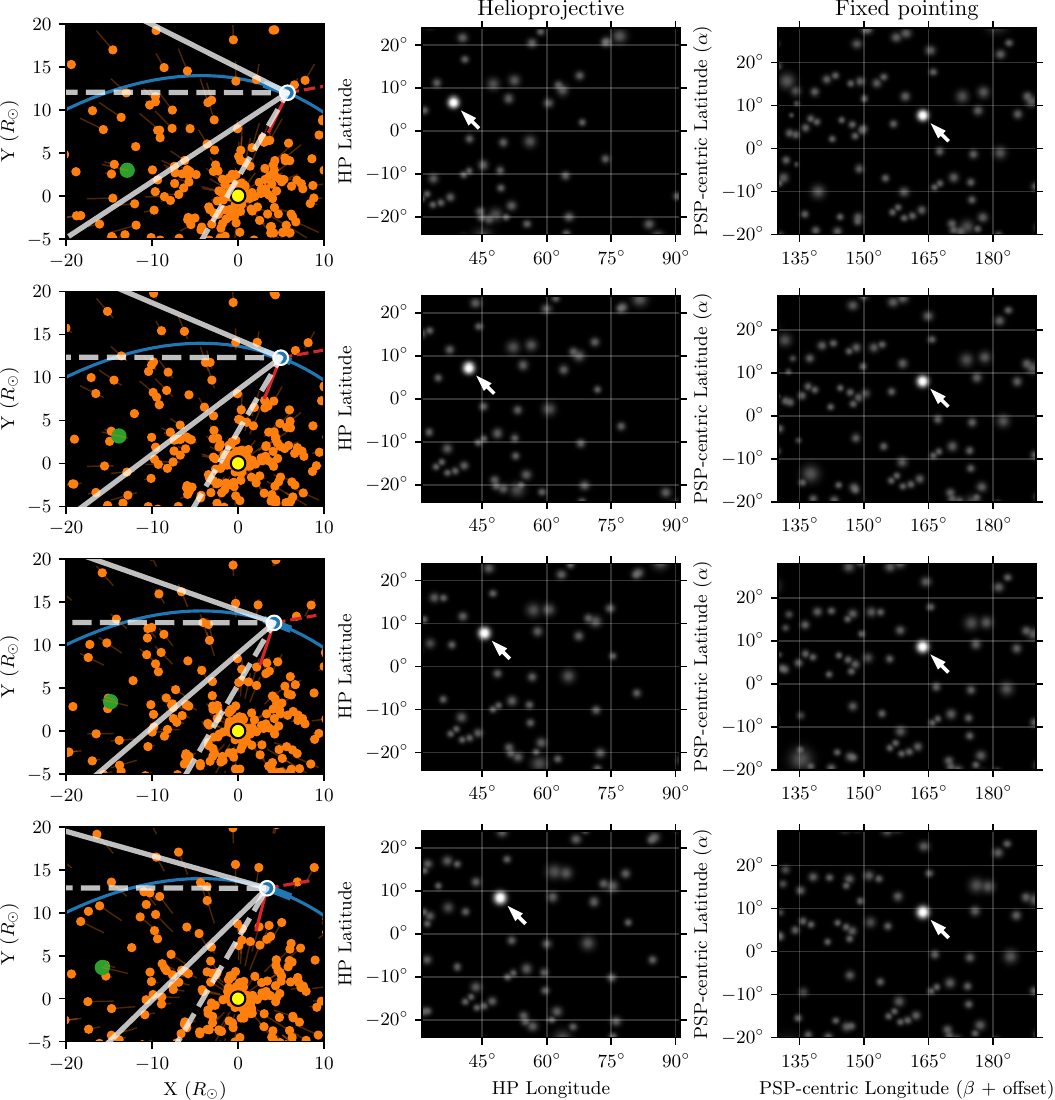}
	\caption{Model demonstration of the stationary point in 3D. The left-hand
	column provides an overhead view at four points in time across a 3-hr
	window. As in Figures~\ref{fig:jmap-model-straight-line-overhead} and
	\ref{fig:jmap-model-psp-overhead}, the blue dot and curve are the spacecraft
	and its trajectory, orange marks plasma parcels, and yellow marks the Sun.
	The solid white lines indicate a camera
	field-of-view with constant helioprojective pointing (which tracks the Sun's
	location as WISPR does), while the dashed white
	lines indicate a field-of-view with a fixed direction from the spacecraft
	location. Red lines in the same styles mark the zero point for longitude in
	each frame. The center column shows images synthesized with the constant
	helioprojective pointing, and the right-hand column shows images with
	fixed-direction pointing. One parcel of interest is at the stationary point.
	It is indicated in the synthesized images by an arrow and by artificially
	increasing its brightness, and in the overhead plot by a green dot. The left
	edge of each synthesized image is the sunward side. (An
	animated version of this figure is available.)}
	\label{fig:3d-model-setup}
\end{figure*}

We again demonstrate this method with our model, using the setup shown in
Figure~\ref{fig:3d-model-setup} containing again realistic PSP motion, but now
launching parcels outside the orbital plane. This means features of interest
will move throughout the image plane, so a 2D time--distance plot cannot capture
these features' motion. Instead we work with a sequence of full images. As in
Section~\ref{sec:curved}, the rotation of the spacecraft must be accounted for.
With the model we synthesize images with a fixed camera pointing. With data, we
would reproject each image into a fixed-camera-pointing frame, thereby
``de-rotating'' the images. (This is discussed further in
Section~\ref{sec:data}.) In both cases we align the axes of the fixed-pointing
pixel grid with the orbital plane, so that horizontal pixel position is related
to $\varepsilon$ and $\beta$, and vertical pixel position is proportional to
$\alpha$. In the fixed-pointing or de-rotated images, we measure longitude
relative to an arbitrary, fixed reference direction, meaning that it can be
converted to $\beta$ by subtracting an offset determined by direction of
$\vec{v}_\textrm{psp}$. That offset is constant as long as the direction of
$\vec{v}_\textrm{psp}$ is well-approximated as constant (which it is in this
model case). The elongation $\varepsilon$ of the in-plane projection of the
parcel is the fixed-pointing longitude minus the longitude of the Sun (which
varies more rapidly than the offset to compute $\beta$).

In the synthesized images of Figure~\ref{fig:3d-model-setup}, it can be seen
that a particular plasma parcel appears at a fixed longitude in the
fixed-pointing images---this is a parcel at the 3D stationary point. Over the
3-hr window, in which all velocities can be approximated as constant, the other
parcels can be seen to drift in longitude (which is more clear in the animated
figure), whereas the stationary point parcel varies only in latitude (i.e.
$\alpha$). In the helioprojective images, in which the camera is pointed
relative to the Sun as with WISPR, all features are seen to move through the
image plane---a motion driven largely by the rotating camera field of view
(which matches the behavior of WISPR).

We click on the longitudinally-stationary parcel in 50 frames evenly
spaced across our 3~hr time window, a duration chosen to match the real-data
demonstraction in Section~\ref{sec:data} and which allows averaging over a
number of angular measurements, and we calculate average values for $\beta$
(and therefore $\varepsilon$) and $\alpha$, and we fit a line to the time series
$\alpha$ values to produce $d\alpha/dt$. For the time-varying quantities
$\varepsilon$ and $\alpha$, these average values will represent the center of
our 3~hr window, at which we consider our solution to be most valid.  We compare
these values to ``ideal'' values produced by selecting the pixel of maximum
intensity in each image, and we find that clicking across many images produces
average values that are accurate to within 1\%. These measured values produce
the constraints shown in Figure~\ref{fig:3d-model-intersects}. From the
intersection point we produce values that very closely match the input values of
the model setup: the actual $\Delta\phi$ of 98\deg\ is recovered as 97.5\deg,
the actual $v_p$ of 194.9~\kms\ is recovered as 192.7~\kms, the actual $\theta$
of 12\deg\ is recovered as 11.97\deg, and the actual $r_p$ of 15~\Rsun\ is
recovered as 15.2~\Rsun. (The $\Delta\phi$ and $r_p$ values refer to the center
of the 3~hr time window, and the other values are constant with time.) This
demonstrates that this method can very closely recover the true values.

In addition to this case study, we have also run this full process in an
automated fashion over a grid of possible parcels placed to appear at the
stationary point, covering $5\deg < \Delta\phi < 130\deg$, $0\deg < \theta <
75\deg$, and $5\,\,\Rsun < r_\textrm{pxy} < 15\,\,\Rsun$ (with a corresponding
$v_p$ computed to place the parcel at the stationary point). The method
successfully recovered the input parameters with similar accuracy across the
grid. Some outlier parcels did not result in successful parameter recovery, but
those were cases that, for instance, required a parcel to move radially
\emph{inward} in order to appear at the stationary point, or to travel at
unrealistic speeds ($> 1000$~\kms), outside the bounds of our numerical solution
for the second constraint.

\section{Application to WISPR images}
\label{sec:data}

\begin{figure*}
	\centering
	\includegraphics{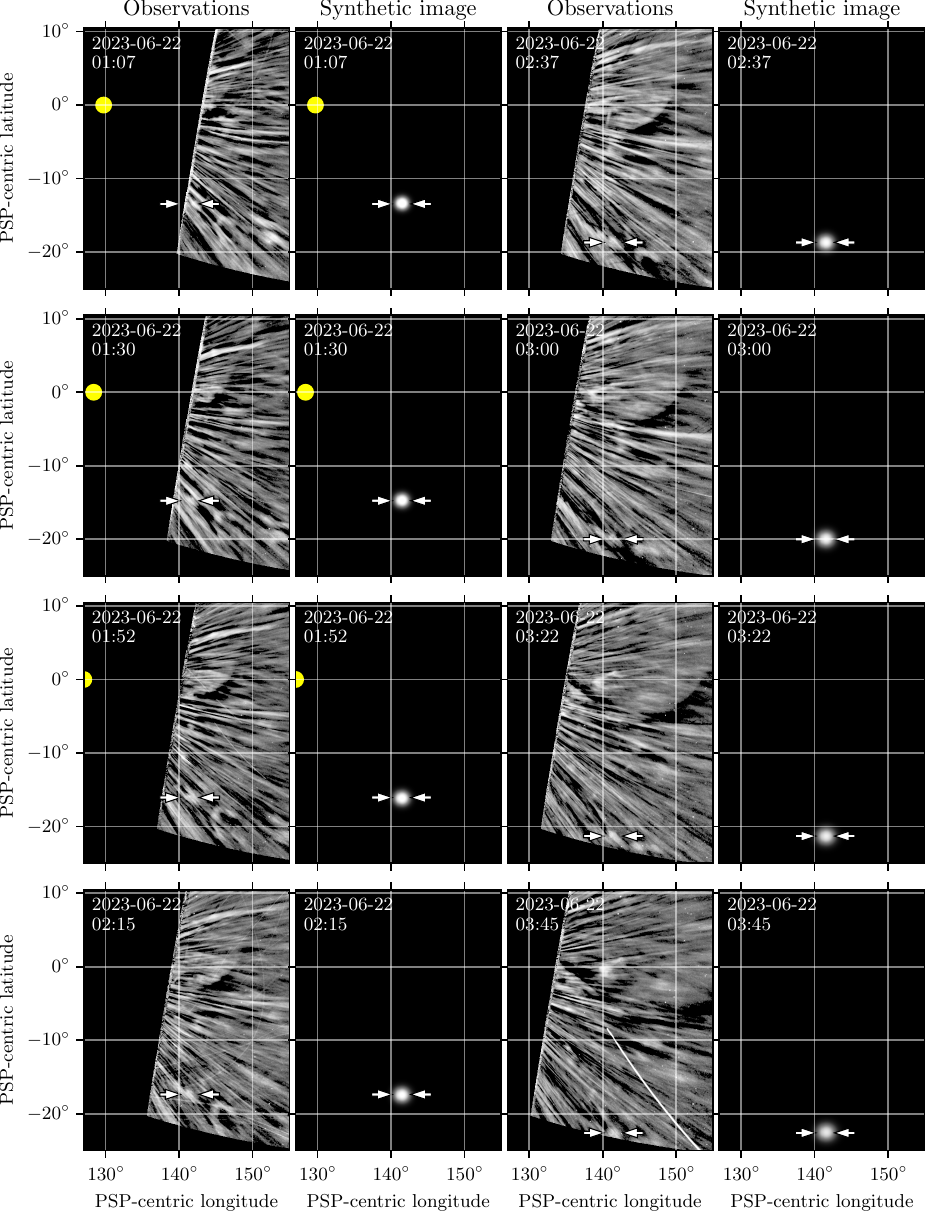}
	\caption{A sequence of images spanning 2.5~hr containing our feature of
	interest, marked by arrows. The first and third columns show LW-processed
	WISPR images, and the other columns show the feature reproduced by our
	forward model and the inferred parameters. The coordinate frame, described
	in the text, is identical for all images, and the arrows are plotted in the
	same location across each pair of real and synthesized images. The yellow
	dot marks the location of the Sun (not to scale), which is a fixed distance
	from the edge of the data and so quickly moves out of frame. (An animated
	version of this figure is available.)}
	\label{fig:data-marked-feature}
\end{figure*}

We now demonstrate the viability of this analysis on images from the WISPR
imager \citep{Vourlidas2016} on PSP. WISPR consists of a pair of white-light
imagers, pointed in PSP's direction of motion with a composite field of view
covering a range from 13.5\deg\ to 108\deg\ from the Sun and approximately
50\deg\ in the transverse direction. PSP travels on a highly eccentric orbit,
with nominal data collection occurring while PSP is below 0.25~au on each orbit.
For the data we use, from orbit 16, PSP's perihelion distance was 13.3~\Rsun. We
use approximately three hours of observations near perihelion. During this
time, PSP's velocity rises from 162.4~\kms\ to 163.0~\kms, and its direction of
motion changes by 5\deg, indicating that PSP can be well approximated as having
a constant velocity. We use LW-processed WISPR images---this technique,
described in the Appendix of \citet{Howard2022}, is similar to difference
imaging in that it isolates the time-variable component of an image sequence. LW
processing uses a sliding window in time and a sequence of filtering steps to
estimate the steady background component in any one pixel. Each image is then
divided by this background to display the time-varying component in relative
terms. This method does an excellent job of removing the steady F-corona, as
well as the steady portion of the K-corona, from each WISPR image, leaving
behind the transient plasma we wish to study. (Importantly, the plasma at the
stationary point is stationary in a fixed or rotation-stabilized reference
frame, but the LW processing operates with respect to variability in the
\textit{image plane}, where this plasma is \textit{not} stationary.)

\begin{figure}
	\centering
	\includegraphics{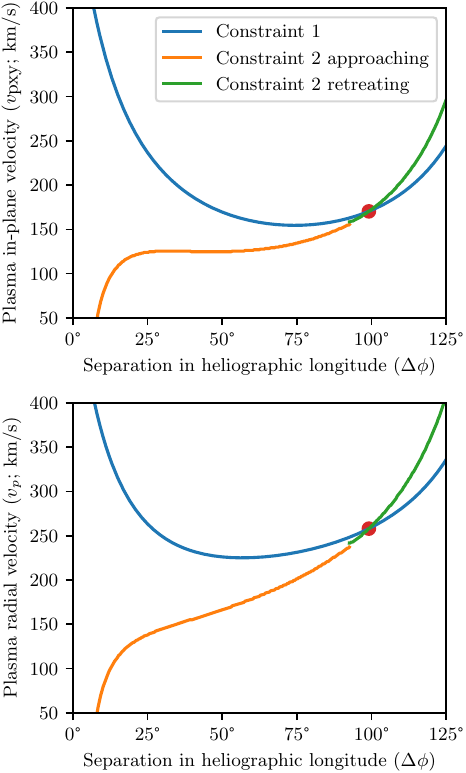}
	\caption{Constraints generated for the parcel marked in
	Figure~\ref{fig:data-marked-feature}. The two panels show in-plane speed
	and the corresponding total speed, respectively, with the conversion between
the two being a function of $\theta$ and therefore $\Delta\phi$.}
	\label{fig:data-constraints}
\end{figure}

The WISPR field of view is very nearly fixed in helioprojective coordinates, or
in other words, it rotates to track the Sun as PSP moves through its orbit.
However, we require stabilized images in an unrotating frame, as discussed in
previous sections. We therefore reproject a sequence of images into an
unrotating frame which we call the ``PSP-centric frame'' (identical to the
``fixed pointing'' frame shown in the right-hand column of
Figure~\ref{fig:3d-model-setup}) in which the PSP-centric longitude represents a
direction from the spacecraft relative to an unchanging reference direction and
PSP-centric latitude is measured relative to PSP's orbital plane\footnote{This
PSP-centric frame is a 3D helioprojective frame (centered on PSP) that is
rotated about the PSP--Sun axis to align its equator with the orbital plane, and
then rotated about its poles so that 0\deg\ longitude is at a fixed but
arbitrary reference direction (for example, toward a chosen distant star),
rather than following the location of the Sun.}. The WISPR images pan through
this reference frame over each encounter.

We identified a sample plasma parcel which appears at a fixed PSP-centric
longitude and increasingly-negative latitude---meaning it is at the 3D
stationary point as described in Section~\ref{sec:3D}. We show a sequence of
frames with this stationary parcel marked in
Figure~\ref{fig:data-marked-feature}. While other features move over it and it
is at times subtle, it appears consistently over 24 frames (of which 8 are shown
in the figure), and it moves with a consistent $d\alpha/dt$ over this 2.5~hr
period while appearing at a fixed longitude.

Over this 2.5~hr sequence, we see the parcel at a longitude of 141\deg, which
corresponds to values of $\varepsilon=15.6\deg$ and $\beta=71.7\deg$ at the
middle of this time window. We also measure $\alpha=-17.4\deg$ (in the middle
frame) and $d\alpha/dt = -3.5\deg$~hr$^{-1}$, following the same procedure as in
Section~\ref{sec:3d-model-demo}. By examining the resulting constraints shown in
Figure~\ref{fig:data-constraints}, we infer the parcel is traveling at a
velocity $v_p=275$~\kms, at an inclination $\theta=-49\deg$ below PSP's orbital
plane. During this window of observations, it is about $\Delta\phi=99\deg$ of
longitude in front of PSP. These angular coordinates correspond to a Carrington
longitude of 228\deg\ and latitude of -49\deg. At the center of the time window,
the parcel is 6.0~\Rsun\ from the Sun and 15.2~\Rsun\ from PSP. Its distance
from the Sun grows from 4.4~\Rsun\ to 7.7~\Rsun\ during the 3~hr window.

Using these inferred parameters, we generate synthetic images of a parcel with
the same trajectory and velocity, using the same forward model of
Section~\ref{sec:method}, which are shown in the second and fourth columns of
Figure~\ref{fig:data-marked-feature}. It can be seen that the position of the
parcel in the images is reproduced very closely, adding credibility to the claim
that this parcel trajectory is implied by the observations, and that the
approximations of constant spacecraft and parcel velocities are reasonable.

\begin{figure*}
	\centering
	\includegraphics{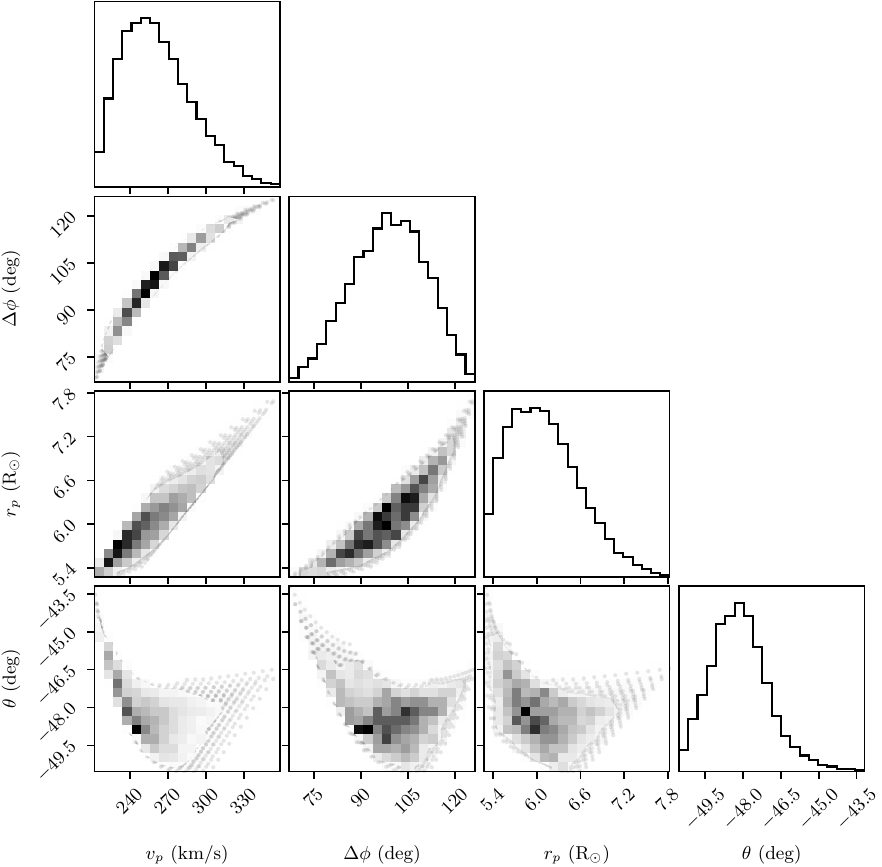}
	\caption{Distributions and correlations of output quantities over a grid of
	assumed errors. See the text for details.}
	\label{fig:corner-plot}
\end{figure*}

Our inferred values are relatively insensitive to measurement error. The
measured $\varepsilon$, $\alpha$ and $d\alpha/dt$ are produced as averages or
fits to marked feature locations in a number of frames, so individual
``mis-clicks'' will tend to balance out. (As shown in Section
\ref{sec:3d-model-demo}, the measured values of these angles are good to within
1\% under ideal conditions.) If we allow an error of $\pm1\deg$ in $\beta$ (and
therefore $\varepsilon$) and $\alpha$ and a $\pm5$\% error in $d\alpha/dt$
(these are generous error margins compared to the spread of values we
find after clicking the feature in each image) and compute the constraints and
implied speed and trajectory across a grid of errors uniformly distributed
within these ranges, we can produce a distribution of possible speeds and
trajectories, which we show in Figure~\ref{fig:corner-plot}. For the
feature we analyze here, the mean and standard deviation of these distributions
are $v_p=262 \pm 27$~\kms, $\theta = -48\deg \pm 1\deg$, $\Delta\phi = 99\deg
\pm 11\deg$, and $r_p = 6.1 \pm 0.5$~\Rsun, showing that the inferred values
carry an uncertainty of only $\sim10$\% due to these assumed ranges of
errors. The correlations that are seen are readily interpreted: the
further away the parcel is placed (in $\Delta\phi$), the further from the Sun it
must be to appear at the same angular location, and the faster it must move to
produce the same angular motion.

\subsection{Comparison to Other Measurements}

A variety of wind speed measurements near the Sun have been reported in the
literature. While there is a range of values, a few hundred \kms\ is
typical. It is useful to compare our measured value of $271 \pm 24$~\kms\ at
6.3~\Rsun. A variety of radio sounding observations \citep[e.g.][and references
therein]{Armstrong1981} span 100--225~\kms\ at 5--10~\Rsun. \citet{DeForest2018}
tracked white-light features seen from Earth and made measurements of
$150-175$~\kms\ at heliocentric distances as low as 7~\Rsun. \citet{Nindos2021}
report measurements of similar density enhancements in the solar wind seen by
WISPR. By assuming all observed features are on the Thompson surface and fitting
the slopes of motion tracks on time-distance plots, they report velocities of
150--300~\kms. In-situ measurements from PSP have recorded wind speeds from
150--500~\kms\ at distances as low as 15~\Rsun\ \citep{Raouafi2023}. From a
theoretical perspective, \citet{Cranmer2019} illustrate that models of wind
acceleration driven by anisotropic magnetohydrodynamic turbulence predict speeds
at 6~\Rsun\ of up to 400~\kms\ for open field lines in coronal hole regions, or
as low as 150~\kms\ in streamers.

When comparing this range of reported values to our result, it is
important to point out that we have measured a single parcel, rather than
producing a distribution of measurements, meaning our result is less robust than
the other values. (Reasons for this limitation and prospects for measuring
additional parcels are discussed in Section~\ref{sec:conclusion}.)
Additionally, given that these reported values will use a variety of averaging
methods and cover different ranges of latitude and levels of solar activity, and
that Earth-bound measurements may be biased toward larger, more visible
features, which may not be representative of the population of features seen by
WISPR, it is difficult to directly compare our single measured parcel to any one
of these studies, but it is encouraging to see that our inferred speed fits
reasonably within the overall range.

\section{Discussion and Conclusions}
\label{sec:conclusion}

In this paper we developed a novel technique for estimating the speeds and
trajectories of certain density enhancements seen by WISPR, though the technique
is applicable to any moving vantage point traveling through a stream of
outflowing objects. WISPR's rapidly-moving vantage point and its close proximity
to the structures being imaged are a hindrance to many flow
tracking methods traditionally applied to the solar wind, but they are the tools
by which this technique works. We showed that the technique can be applied
despite WISPR's sun-tracking rotation and the elliptical orbit on which it
travels. We demonstrated the method by analyzing one plasma parcel seen by
WISPR, producing an estimated speed and trajectory which, in turn, reproduce the
observations.

Our method has several limitations, the largest of which is that only plasma
features which appear at their stationary point can be measured. As shown by
Equation~\ref{eqn:constr1}, the location of the stationary point is a function
of the parcel's speed and its angular distance from the spacecraft, meaning that
in any image sequence there will be a range of possible stationary points at
which parcels could appear, and only those parcels that by coincidence do appear
at that point (meaning that, momentarily, they are on course to ``collide'' with
the spacecraft, or to pass directly above or below it) can be measured. The
technique also can only measure parcels whose stationary point falls within the
field of view. The higher the parcel speed relative to the spacecraft speed, the
more Sun-ward the corresponding stationary point will be, and so the fastest
parcels (which may include all parcels when the spacecraft velocity is low, far
from perihelion) may not be measurable as their stationary point lies outside
the field of view.

Second, we must assume that features seen at the stationary point are
individual, relatively compact density enhancements moving linearly. Possible
alternatives include, for instance, an elongated radial structure with a fold or
perturbation (producing the observed density feature) that is propagating along
the structure, or a coincidence of several unrelated, spatially disjoint, highly
transitory features that appear in turn to produce the appearance of a single
feature at fixed $\beta$ and with a constant $d\alpha/dt$. Such an assumption is
foundational to many wind speed measurements, but it is important to state
specifically.

Third, we require a few assumptions to be valid for the velocities. The
spacecraft velocity must be well-approximated as constant (in both magnitude and
direction) for the period of time being analyzed, and the parcel being observed
must be well-approximated as traveling radially-out from the Sun at a constant
velocity during the time period. In our demonstration in Section~\ref{sec:data},
using a 3~hr window of time, these assumptions are reasonable as discussed in
that section, but this may not always be the case.

Finally, addressing our science goal of constraining the speed, and therefore
acceleration, of the ambient solar wind, we must assume that the discrete
density features we can measure are passive tracers of the ambient wind, rather
than separate, more transient features undergoing distinct physical processes.

Despite these limitations, this is a novel and useful method. The most similar
method that these authors are aware of is that of \citet{Liewer2020}, in which a
feature of interest is identified in a sequence of images, and then a parcel
trajectory and speed is iteratively fit that reproduces the apparent location of
the feature in the images. That method is more widely applicable, as it does not
require the coincidence of the feature of interest appearing at the stationary
point. (Though our limitation to a random subset of plasma parcels that appear
at the stationary point may represent a convenient way of surveying a random
sample of parcels relatively free of human biases.) However, our method is
numerically simpler in that it does not require iterative, non-linear fitting,
and so there is no concern over sensitivity to initial guesses.

In future work, we will explore the possibility of extending our method's
applicability to parcels not at the stationary point by treating parcels'
horizontal angular positions similarly to their vertical positions. We will also
conduct a census of plasma features seen at the stationary point by WISPR and
produce a catalog of inferred velocities, latitudes, etc. We expect that this
will provide a sampling of wind speeds across a range of latitudes, all at close
proximity to the Sun and in the critical early acceleration region.
	
\begin{acknowledgements}

We thank M. G. Linton and P. C. Liewer for reviewing this manuscript and
offering valuable suggestions.
This research has made use of NASA's Astrophysics Data System Bibliographic
Services, which these authors treasure. We specifically acknowledge the
Astropy/SunPy coordinate framework, which has greatly simplified a number of
operations in this work. This work was supported by the NASA Parker Solar Probe
Program Office for the WISPR program (NASA Contract NNG11EK11I to the U.S. Naval
Research Laboratory, NRL; and NRL subcontract N00173-20-C-2002 to SwRI). Parker
Solar Probe was designed, built, and is now operated by the Johns Hopkins
Applied Physics Laboratory as part of NASA's Living with a Star (LWS) program
(contract NNN06AA01C). Support from the LWS management and technical team has
played a critical role in the success of the Parker Solar Probe mission. The
Wide-Field Imager for Parker Solar Probe (WISPR) instrument was designed, built,
and is now operated by the US Naval Research Laboratory in collaboration with
Johns Hopkins University/Applied Physics Laboratory, California Institute of
Technology/Jet Propulsion Laboratory, University of Gottingen, Germany, Centre
Spatiale de Liege, Belgium and University of Toulouse/Research Institute in
Astrophysics and Planetology.

\end{acknowledgements}

\software{SunPy \citep{SunpyCommunity2020} version 5.1.2 \citep{sunpy5.1.2},
Astropy \citep{astropy:2013, astropy:2018, astropy:2022} version 6.1.0
\citep{astropy6.1.0}, Matplotlib \citep{Hunter2007} version 3.3.2
\citep{matplotlib3.8.4}, NumPy \citep{Harris2020} version 1.26.4, SciPy
\citep{Virtanen2020} version 1.13.0 \citep{scipy1.13.0}}

\bibliography{library}{}

\begin{thebibliography}{}
\expandafter\ifx\csname natexlab\endcsname\relax\def\natexlab#1{#1}\fi
\providecommand{\url}[1]{\href{#1}{#1}}
\providecommand{\dodoi}[1]{doi:~\href{http://doi.org/#1}{\nolinkurl{#1}}}
\providecommand{\doeprint}[1]{\href{http://ascl.net/#1}{\nolinkurl{http://ascl.net/#1}}}
\providecommand{\doarXiv}[1]{\href{https://arxiv.org/abs/#1}{\nolinkurl{https://arxiv.org/abs/#1}}}

\bibitem[{Armstrong \& Woo(1981)}]{Armstrong1981}
Armstrong, J.~W., \& Woo, R. 1981, Astronomy and Astrophysics, 103, 415

\bibitem[{{Astropy Collaboration}(2024)}]{astropy6.1.0}
{Astropy Collaboration}. 2024, Astropy 6.1.0, Zenodo,
  \dodoi{10.5281/zenodo.11121433}

\bibitem[{{Astropy Collaboration} {et~al.}(2013){Astropy Collaboration},
  Robitaille, Tollerud, Greenfield, Droettboom, Bray, Aldcroft, Davis,
  Ginsburg, {Price-Whelan}, Kerzendorf, Conley, Crighton, Barbary, Muna,
  Ferguson, Grollier, Parikh, Nair, Unther, Deil, Woillez, Conseil, Kramer,
  Turner, Singer, Fox, Weaver, Zabalza, Edwards, Azalee~Bostroem, Burke, Casey,
  Crawford, Dencheva, Ely, Jenness, Labrie, Lian~Lim, Pierfederici, Pontzen,
  Ptak, Refsdal, Servillat, \& Streicher}]{astropy:2013}
{Astropy Collaboration}, Robitaille, T.~P., Tollerud, E.~J., {et~al.} 2013,
  558, A33, \dodoi{10.1051/0004-6361/201322068}

\bibitem[{{Astropy Collaboration} {et~al.}(2018){Astropy Collaboration},
  {Price-Whelan}, Sip{\H o}cz, G{\"u}nther, Lim, Crawford, Conseil, Shupe,
  Craig, Dencheva, Ginsburg, VanderPlas, Bradley, {P{\'e}rez-Su{\'a}rez}, {de
  Val-Borro}, Paper~Contributors, Aldcroft, Cruz, Robitaille, Tollerud,
  Coordination~Committee, Ardelean, Babej, Bach, Bachetti, Bakanov, Bamford,
  Barentsen, Barmby, Baumbach, Berry, Biscani, Boquien, Bostroem, Bouma,
  Brammer, Bray, Breytenbach, Buddelmeijer, Burke, Calderone,
  Cano~Rodr{\'{\i}}guez, Cara, Cardoso, Cheedella, Copin, Corrales, Crichton,
  D{\'A}vella, Deil, Depagne, Dietrich, Donath, Droettboom, Earl, Erben,
  Fabbro, Ferreira, Finethy, Fox, Garrison, Gibbons, Goldstein, Gommers, Greco,
  Greenfield, Groener, Grollier, Hagen, Hirst, Homeier, Horton, Hosseinzadeh,
  Hu, Hunkeler, Ivezi{\'c}, Jain, Jenness, Kanarek, Kendrew, Kern, Kerzendorf,
  Khvalko, King, Kirkby, Kulkarni, Kumar, Lee, Lenz, Littlefair, Ma, Macleod,
  Mastropietro, McCully, Montagnac, Morris, Mueller, Mumford, Muna, Murphy,
  Nelson, Nguyen, Ninan, N{\"o}the, Ogaz, Oh, Parejko, Parley, Pascual, Patil,
  Patil, Plunkett, Prochaska, Rastogi, Reddy~Janga, Sabater, Sakurikar,
  Seifert, Sherbert, {Sherwood-Taylor}, Shih, Sick, Silbiger, Singanamalla,
  Singer, Sladen, Sooley, Sornarajah, Streicher, Teuben, Thomas, Tremblay,
  Turner, Terr{\'o}n, {van Kerkwijk}, {de la Vega}, Watkins, Weaver, Whitmore,
  Woillez, Zabalza, \& Contributors}]{astropy:2018}
{Astropy Collaboration}, {Price-Whelan}, A.~M., Sip{\H o}cz, B.~M., {et~al.}
  2018, 156, 123, \dodoi{10.3847/1538-3881/aabc4f}

\bibitem[{{Astropy Collaboration} {et~al.}(2022){Astropy Collaboration},
  {Price-Whelan}, Lim, Earl, Starkman, Bradley, Shupe, Patil, Corrales,
  Brasseur, N{\"o}the, Donath, Tollerud, Morris, Ginsburg, Vaher, Weaver,
  Tocknell, Jamieson, {van Kerkwijk}, Robitaille, Merry, Bachetti, G{\"u}nther,
  Aldcroft, {Alvarado-Montes}, Archibald, B{\'o}di, Bapat, Barentsen,
  Baz{\'a}n, Biswas, Boquien, Burke, Cara, Cara, Conroy, Conseil, Craig, Cross,
  Cruz, D'Eugenio, Dencheva, Devillepoix, Dietrich, Eigenbrot, Erben, Ferreira,
  {Foreman-Mackey}, Fox, Freij, Garg, Geda, Glattly, Gondhalekar, Gordon,
  Grant, Greenfield, Groener, Guest, Gurovich, Handberg, Hart,
  {Hatfield-Dodds}, Homeier, Hosseinzadeh, Jenness, Jones, Joseph, Kalmbach,
  Karamehmetoglu, Ka{\l}uszy{\'n}ski, Kelley, Kern, Kerzendorf, Koch, Kulumani,
  Lee, Ly, Ma, MacBride, Maljaars, Muna, Murphy, Norman, O'Steen, Oman,
  Pacifici, Pascual, {Pascual-Granado}, Patil, Perren, Pickering, Rastogi,
  Roulston, Ryan, Rykoff, Sabater, Sakurikar, Salgado, Sanghi, Saunders,
  Savchenko, Schwardt, {Seifert-Eckert}, Shih, Jain, Shukla, Sick, Simpson,
  Singanamalla, Singer, Singhal, Sinha, Sip{\H o}cz, Spitler, Stansby,
  Streicher, {\v S}umak, Swinbank, Taranu, Tewary, Tremblay, de~{Val-Borro},
  Van~Kooten, Vasovi{\'c}, Verma, {de Miranda Cardoso}, Williams, Wilson,
  Winkel, {Wood-Vasey}, Xue, Yoachim, Zhang, Zonca, \& {Astropy Project
  Contributors}}]{astropy:2022}
{Astropy Collaboration}, {Price-Whelan}, A.~M., Lim, P.~L., {et~al.} 2022, 935,
  167, \dodoi{10.3847/1538-4357/ac7c74}

\bibitem[{Bale {et~al.}(2023)Bale, Drake, McManus, Desai, Badman, Larson,
  Swisdak, Horbury, Raouafi, Phan, Velli, McComas, Cohen, Mitchell, Panasenco,
  \& Kasper}]{Bale2023}
Bale, S.~D., Drake, J.~F., McManus, M.~D., {et~al.} 2023, Nature, 618, 252,
  \dodoi{10.1038/s41586-023-05955-3}

\bibitem[{Braga \& Vourlidas(2021)}]{Braga2021}
Braga, C.~R., \& Vourlidas, A. 2021, Astronomy \& Astrophysics, 650, A31,
  \dodoi{10.1051/0004-6361/202039490}

\bibitem[{Cranmer {et~al.}(2017)Cranmer, Gibson, \& Riley}]{Cranmer2017}
Cranmer, S.~R., Gibson, S.~E., \& Riley, P. 2017, Space Science Reviews, 212,
  1345, \dodoi{10.1007/s11214-017-0416-y}

\bibitem[{Cranmer \& Winebarger(2019)}]{Cranmer2019}
Cranmer, S.~R., \& Winebarger, A.~R. 2019, Annual Review of Astronomy and
  Astrophysics, 57, 157, \dodoi{10.1146/annurev-astro-091918-104416}

\bibitem[{DeForest {et~al.}(2018)DeForest, Howard, Velli, Viall, \&
  Vourlidas}]{DeForest2018}
DeForest, C.~E., Howard, R.~A., Velli, M., Viall, N., \& Vourlidas, A. 2018,
  The Astrophysical Journal, 862, 18, \dodoi{10.3847/1538-4357/AAC8E3}

\bibitem[{Fox {et~al.}(2016)Fox, Velli, Bale, Decker, Driesman, Howard, Kasper,
  Kinnison, Kusterer, Lario, Lockwood, McComas, Raouafi, \& Szabo}]{Fox2016}
Fox, N.~J., Velli, M.~C., Bale, S.~D., {et~al.} 2016, Space Science Reviews,
  204, 7, \dodoi{10.1007/s11214-015-0211-6}

\bibitem[{Gommers {et~al.}(2024)Gommers, Virtanen, Haberland, Burovski,
  Weckesser, Reddy, Oliphant, Cournapeau, Nelson, {alexbrc}, Roy, Peterson,
  Polat, Wilson, {endolith}, Mayorov, van~der Walt, Brett, Laxalde, Larson,
  Sakai, Millman, Lars, {peterbell10}, Carey, van Mulbregt, {eric-jones},
  McKibben, Kai, \& Kern}]{scipy1.13.0}
Gommers, R., Virtanen, P., Haberland, M., {et~al.} 2024, Scipy/Scipy: {{SciPy}}
  1.13.0, Zenodo, \dodoi{10.5281/zenodo.10909890}

\bibitem[{Harris {et~al.}(2020)Harris, Millman, {van der Walt}, Gommers,
  Virtanen, Cournapeau, Wieser, Taylor, Berg, Smith, Kern, Picus, Hoyer, {van
  Kerkwijk}, Brett, Haldane, {del R{\'i}o}, Wiebe, Peterson,
  {G{\'e}rard-Marchant}, Sheppard, Reddy, Weckesser, Abbasi, Gohlke, \&
  Oliphant}]{Harris2020}
Harris, C.~R., Millman, K.~J., {van der Walt}, S.~J., {et~al.} 2020, Nature,
  585, 357, \dodoi{10.1038/s41586-020-2649-2}

\bibitem[{Howard {et~al.}(2022)Howard, Stenborg, Vourlidas, Gallagher, Linton,
  Hess, Rich, \& Liewer}]{Howard2022}
Howard, R.~A., Stenborg, G., Vourlidas, A., {et~al.} 2022, The Astrophysical
  Journal, 936, 43, \dodoi{10.3847/1538-4357/ac7ff5}

\bibitem[{Hunter(2007)}]{Hunter2007}
Hunter, J.~D. 2007, Computing in Science and Engineering, 9, 90,
  \dodoi{10.1109/MCSE.2007.55}

\bibitem[{Kenny {et~al.}(2023)Kenny, DeForest, West, \& Liewer}]{Kenny2023}
Kenny, M.~N., DeForest, C.~E., West, M.~J., \& Liewer, P.~C. 2023, The
  Astrophysical Journal, 953, 79, \dodoi{10.3847/1538-4357/acdfc5}

\bibitem[{Liewer {et~al.}(2019)Liewer, Vourlidas, Thernisien, Qiu, Penteado,
  Nistic{\`o}, Howard, \& Bothmer}]{Liewer2019}
Liewer, P., Vourlidas, A., Thernisien, A., {et~al.} 2019, Solar Physics, 294,
  93, \dodoi{10.1007/s11207-019-1489-4}

\bibitem[{Liewer {et~al.}(2022)Liewer, Qiu, Ark, Penteado, Stenborg, Vourlidas,
  Hall, \& Riley}]{Liewer2022}
Liewer, P.~C., Qiu, J., Ark, F., {et~al.} 2022, Solar Physics, 297, 128,
  \dodoi{10.1007/s11207-022-02058-6}

\bibitem[{Liewer {et~al.}(2020)Liewer, Qiu, Penteado, Hall, Vourlidas, \&
  Howard}]{Liewer2020}
Liewer, P.~C., Qiu, J., Penteado, P., {et~al.} 2020, Solar Physics, 295, 1,
  \dodoi{10.1007/s11207-020-01715-y}

\bibitem[{Liewer {et~al.}(2021)Liewer, Qiu, Vourlidas, Hall, \&
  Penteado}]{Liewer2021}
Liewer, P.~C., Qiu, J., Vourlidas, A., Hall, J.~R., \& Penteado, P. 2021,
  Astronomy \& Astrophysics, 650, A32, \dodoi{10.1051/0004-6361/202039641}

\bibitem[{Liewer {et~al.}(2023)Liewer, Vourlidas, Stenborg, Howard, Qiu,
  Penteado, Panasenco, \& Braga}]{Liewer2023}
Liewer, P.~C., Vourlidas, A., Stenborg, G., {et~al.} 2023, The Astrophysical
  Journal, 948, 24, \dodoi{10.3847/1538-4357/acc8c7}

\bibitem[{{Matplotlib Development Team}(2024)}]{matplotlib3.8.4}
{Matplotlib Development Team}. 2024, Matplotlib 3.8.4, Zenodo,
  \dodoi{10.5281/zenodo.10916799}

\bibitem[{Mumford {et~al.}(2024)Mumford, Freij, Stansby, Christe, Ireland,
  Shih, Mayer, Hughitt, Ryan, Liedtke, Hayes, Barnes, {P{\'e}rez-Su{\'a}rez},
  I, Chakraborty, Inglis, Pattnaik, Sip{\H o}cz, MacBride, Sharma, Leonard,
  Hewett, Hamilton, Manhas, Panda, Earnshaw, Choudhary, Kumar, Singh, Chanda,
  Haque, Kirk, Mueller, Konge, {Wentzel-Long}, Srivastava, Maloney, Jain,
  Zivadinovic, Bennett, Wilson, Baruah, Arbolante, Simon, Charlton, Mishra,
  Paul, Verma, Chorley, Chouhan, Himanshu, Mason, Modi, Sharma, Naman9639,
  Bobra, Tyagi, Rozo, Manley, Ivashkiv, Laitinen, Chatterjee, von Forstner,
  Baz{\'a}n, Stern, Shukla, Gieseler, Evans, Jain, Malocha, Ghosh,
  Airmansmith97, Sta{\'n}czak, Singh, Visscher, Verma, SophieLemos, Agrawal,
  Alam, Buddhika, Collier, Pathak, Rideout, Sharma, Park, Bates, {ryuusama},
  Shukla, Giger, Mishra, Sharma, Goel, Taylor, Cetusic, Reiter, Jacob,
  Inchaurrandieta, Dacie, Dubey, Eigenbrot, Bray, Murphy, Surve, Zahniy, Sidhu,
  Meszaros, Parkhi, Russell, Bose, Pandey, {Price-Whelan}, J, Chicrala, Ankit,
  Graham, Guennou, D'Avella, Williams, Verma, Ballew, Agrawal, Singh, Lodha,
  Robitaille, Augspurger, Krishan, {honey}, {neerajkulk}, Bhope, Gaba, Hill,
  Dixit, Mampaey, Wiedemann, Molina, Briseno, Ke{\c s}kek, Habib, Letts,
  Singaravelan, Ranjan, Altunian, Streicher, Gomillion, Agarwal, Kothari,
  Nomiya, {mridulpandey}, Stevens, B, Bahuleyan, Shauryam, Kaszynski, W,
  Mehrotra, Tang, Sinha, Smith, Sinha, Kustov, Stone, Bard, Behn, Arias,
  Tollerud, Dover, Verstringe, Kumar, Mathur, Babuschkin, Calixto, Wimbish,
  Qing, {Buitrago-Casas}, Krishna, Chaudhari, Hiware, Ghosh, McKee, Lyes,
  Mangaonkar, Cheung, Mendero, Dedhia, Schoentgen, Shahdadpuri, Srinivasan,
  Gyenge, OussCHE, Wright, Mekala, Das, Mishra, Sharma, Badman, Kooten,
  Srikanth, Jain, Farah, Kannojia, Chistie, Qing, Yadav, Paul, Wilkinson,
  Caswell, Braccia, Pereira, Gates, Dang, Bankar, Jamieson, Agrawal, {ejm4567},
  {platipo}, {pradeep}, {resakra}, {tal66}, {yasintoda}, Attie, \&
  Murray}]{sunpy5.1.2}
Mumford, S.~J., Freij, N., Stansby, D., {et~al.} 2024, {{SunPy}} 5.1.2, Zenodo,
  \dodoi{10.5281/zenodo.10927245}

\bibitem[{Nindos {et~al.}(2021)Nindos, Patsourakos, Vourlidas, Liewer,
  Penteado, \& Hall}]{Nindos2021}
Nindos, A., Patsourakos, S., Vourlidas, A., {et~al.} 2021, Astronomy \&
  Astrophysics, 650, A30, \dodoi{10.1051/0004-6361/202039414}

\bibitem[{Nistic{\`o} {et~al.}(2020)Nistic{\`o}, Bothmer, Vourlidas, Liewer,
  Thernisien, Stenborg, \& Howard}]{Nistico2020}
Nistic{\`o}, G., Bothmer, V., Vourlidas, A., {et~al.} 2020, Solar Physics, 295,
  63, \dodoi{10.1007/s11207-020-01626-y}

\bibitem[{Raouafi {et~al.}(2023)Raouafi, Stenborg, Seaton, Wang, Wang,
  DeForest, Bale, Drake, Uritsky, Karpen, DeVore, Sterling, Horbury, Harra,
  Bourouaine, Kasper, Kumar, Phan, \& Velli}]{Raouafi2023}
Raouafi, N.~E., Stenborg, G., Seaton, D.~B., {et~al.} 2023, The Astrophysical
  Journal, 945, 28, \dodoi{10.3847/1538-4357/acaf6c}

\bibitem[{Sheeley {et~al.}(1997)Sheeley, Wang, Hawley, Brueckner, Dere, Howard,
  Koomen, Korendyke, Michels, Paswaters, Socker, St.~Cyr, Wang, Lamy, Llebaria,
  Schwenn, Simnett, Plunkett, \& Biesecker}]{Sheeley1997}
Sheeley, N.~R., Wang, Y.~M., Hawley, S.~H., {et~al.} 1997, The Astrophysical
  Journal, 484, 472, \dodoi{10.1086/304338}

\bibitem[{{SunPy Community} {et~al.}(2020){SunPy Community}, Barnes, Bobra,
  Christe, Freij, Hayes, Ireland, Mumford, {Perez-Suarez}, Ryan, Shih,
  Contributors), Chanda, Glogowski, Hewett, Hughitt, Hill, Hiware, Inglis,
  Kirk, Konge, Mason, Maloney, Murray, Panda, Park, Pereira, Reardon, Savage,
  Sip{\H o}cz, Stansby, Jain, Taylor, Yadav, {Rajul}, Dang, \&
  Contributors)}]{SunpyCommunity2020}
{SunPy Community}, Barnes, W.~T., Bobra, M.~G., {et~al.} 2020, The
  Astrophysical Journal, 890, 68, \dodoi{10.3847/1538-4357/ab4f7a}

\bibitem[{Thompson(2009)}]{Thompson2009}
Thompson, W.~T. 2009, Icarus, 200, 351, \dodoi{10.1016/j.icarus.2008.12.011}

\bibitem[{Viall \& Borovsky(2020)}]{Viall2020}
Viall, N.~M., \& Borovsky, J.~E. 2020, Journal of Geophysical Research: Space
  Physics, 125, e2018JA026005, \dodoi{10.1029/2018JA026005}

\bibitem[{Viall \& Vourlidas(2015)}]{Viall2015}
Viall, N.~M., \& Vourlidas, A. 2015, The Astrophysical Journal, 807, 176,
  \dodoi{10.1088/0004-637X/807/2/176}

\bibitem[{Virtanen {et~al.}(2020)Virtanen, Gommers, Oliphant, Haberland, Reddy,
  Cournapeau, Burovski, Peterson, Weckesser, Bright, {van der Walt}, Brett,
  Wilson, Millman, Mayorov, Nelson, Jones, Kern, Larson, Carey, Polat, Feng,
  Moore, VanderPlas, Laxalde, Perktold, Cimrman, Henriksen, Quintero, Harris,
  Archibald, Ribeiro, Pedregosa, {van Mulbregt}, Vijaykumar, Bardelli,
  Rothberg, Hilboll, Kloeckner, Scopatz, Lee, Rokem, Woods, Fulton, Masson,
  H{\"a}ggstr{\"o}m, Fitzgerald, Nicholson, Hagen, Pasechnik, Olivetti, Martin,
  Wieser, Silva, Lenders, Wilhelm, Young, Price, Ingold, Allen, Lee, Audren,
  Probst, Dietrich, Silterra, Webber, Slavi{\v c}, Nothman, Buchner, Kulick,
  Sch{\"o}nberger, {de Miranda Cardoso}, Reimer, Harrington, Rodr{\'i}guez,
  {Nunez-Iglesias}, Kuczynski, Tritz, Thoma, Newville, K{\"u}mmerer,
  Bolingbroke, Tartre, Pak, Smith, Nowaczyk, Shebanov, Pavlyk, Brodtkorb, Lee,
  McGibbon, Feldbauer, Lewis, Tygier, Sievert, Vigna, Peterson, More, Pudlik,
  Oshima, Pingel, Robitaille, Spura, Jones, Cera, Leslie, Zito, Krauss,
  Upadhyay, Halchenko, \& {V{\'a}zquez-Baeza}}]{Virtanen2020}
Virtanen, P., Gommers, R., Oliphant, T.~E., {et~al.} 2020, Nature Methods, 17,
  261, \dodoi{10.1038/s41592-019-0686-2}

\bibitem[{Vourlidas {et~al.}(2016)Vourlidas, Howard, Plunkett, Korendyke,
  Thernisien, Wang, Rich, Carter, Chua, Socker, Linton, Morrill, Lynch, Thurn,
  Van~Duyne, Hagood, Clifford, Grey, Velli, Liewer, Hall, DeJong, Mikic,
  Rochus, Mazy, Bothmer, \& Rodmann}]{Vourlidas2016}
Vourlidas, A., Howard, R.~A., Plunkett, S.~P., {et~al.} 2016, Space Science
  Reviews, 204, 83, \dodoi{10.1007/s11214-014-0114-y}

\bibitem[{Wexler {et~al.}(2020)Wexler, Imamura, Efimov, Song, Lukanina, Ando,
  Jensen, Vierinen, \& Coster}]{Wexler2020}
Wexler, D., Imamura, T., Efimov, A., {et~al.} 2020, Solar Physics, 295, 111,
  \dodoi{10.1007/s11207-020-01677-1}

\end{thebibliography}
\bibliographystyle{aasjournal}

\end{document}